\documentclass[10pt]{article}
\usepackage[utf8]{inputenc}
\usepackage{wrapfig}

\usepackage[margin=0.8in]{geometry}

\usepackage{lipsum}

\usepackage[ruled,vlined]{algorithm2e}
\usepackage{algorithmic}

\newcommand\blfootnote[1]{%
  \begingroup
  \renewcommand\thefootnote{}\footnote{#1}%
  \addtocounter{footnote}{-1}%
  \endgroup
}
\author{Hyeji Kim\blfootnote{H. Kim is with the Department of Electrical and Computer Engineering at University of Texas at Austin. Y. Jiang and S. Kannan are with the Department of Electrical Engineering at University of Washington. S. Oh is with the Department of Computer Science and Engineering at University of Washington. P. Viswanath is with the Department of Electrical Engineering at University of Illinois at Urbana Champaign. 
}, Yihan Jiang, Sreeram Kannan, Sewoong Oh, Pramod Viswanath}

\usepackage{cite}

\usepackage{graphicx}
\usepackage{xcolor}
\usepackage{url}
\usepackage{subcaption}
\usepackage{sectsty,amsmath}
\usepackage{hyperref}

\title{Deepcode and Modulo-SK are Designed for Different Settings}
\date{}

\begin{document}

\maketitle



\begin{abstract}
We respond to \cite{shayevitz2018} which claimed that ``Modulo-SK scheme outperforms Deepcode \cite{kim2018deepcode}". We demonstrate that this statement is {\em not true}: the two schemes are designed and  evaluated for {\em entirely different} settings: DeepCode is designed  and evaluated for the AWGN channel with (potentially delayed) uncoded output feedback. Modulo-SK is evaluated on the AWGN channel with  coded feedback {\color{black} and unit delay}.  \cite{shayevitz2018} also claimed an implementation of Schalkwijk and Kailath (SK)\cite{schalkwijk1966coding}  which was numerically stable for any number of information bits and iterations. {\color{black} 
However, we observe that while their implementation does marginally improve over ours, it also suffers from a fundamental issue with precision.} Finally, we show that Deepcode dominates the optimized performance of SK, over a natural choice of parameterizations when the feedback is noisy. 
\end{abstract}

\section{Deepcode and Modulo-SK are designed for, and work in, different settings.}\label{sec1}
Deepcode is designed for channels with passive (delayed) output feedback. Modulo-SK is designed for channels with active feedback with {\color{black} unit delay}. The difference is whether the receiver is allowed to \emph{encode} the feedback signal and how much delay there is before this feedback reaches the transmitter.

By passive feedback, we refer to the setting where the decoder cannot encode the feedback sent to the transmitter. The active feedback setting allows for this additional degree of freedom at the decoder. We note that the difference between active and passive settings has been well noted in the literature\cite{kim2007gaussian,kim2011error}. For example, in the passive setting with noisy feedback, ``linear coding schemes incorporating noisy feedback fail to achieve any positive rate'' \cite{kim2007gaussian}, highlighting the hardness of this setting. Indeed, this shortcoming in the passive feedback setting was precisely the motivation and setting for our paper. 



\begin{figure}[!ht]
  \centering
  \begin{subfigure}[b]{0.45\linewidth}
    \centering\includegraphics[width=\textwidth]{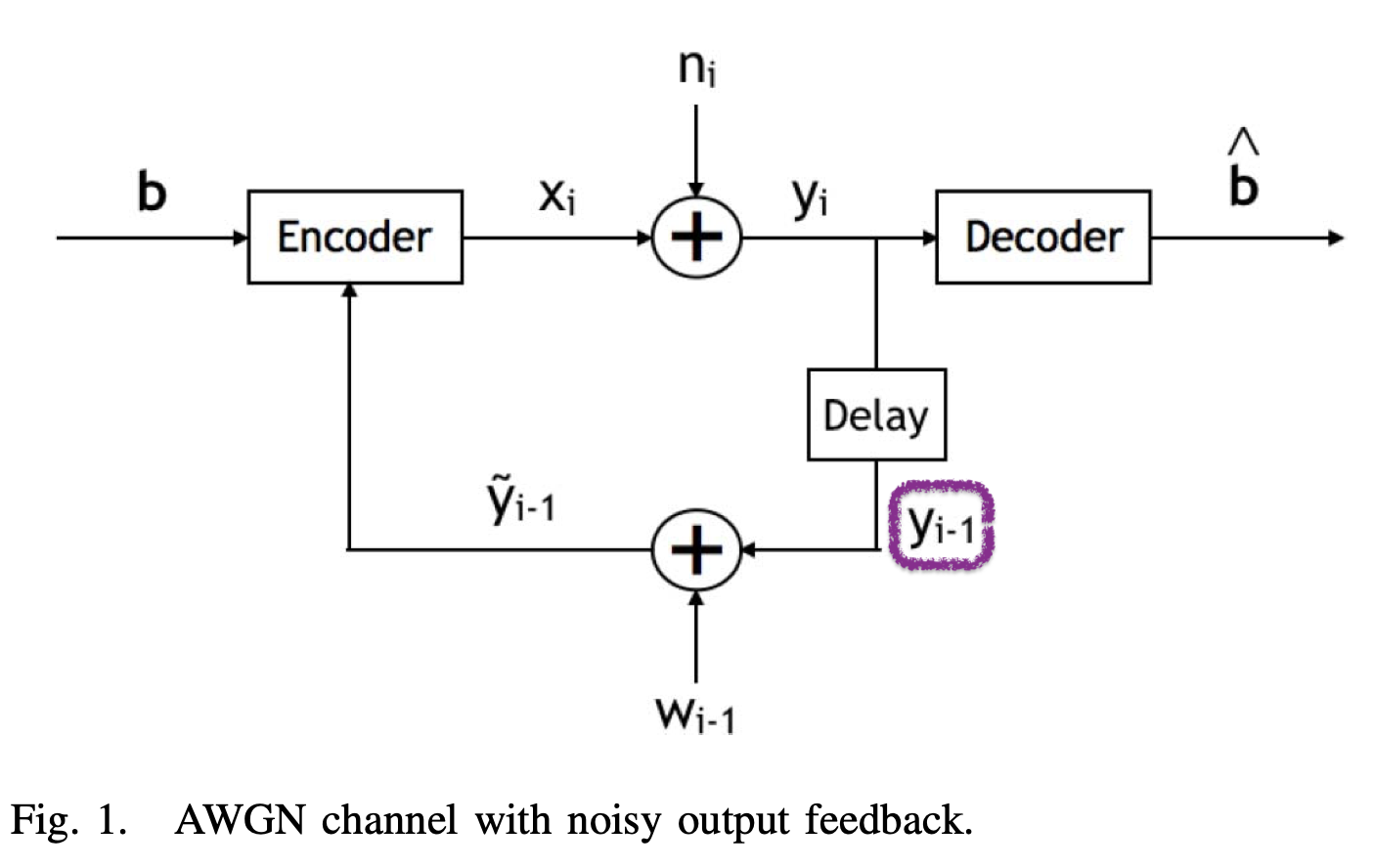}
    \caption{\label{fig:fig1} Figure 1 of the Deepcode papers~\cite{kim2018deepcode,deepcode2018jsait}}
  \end{subfigure}%
  \begin{subfigure}[b]{0.57\linewidth}
    \centering\includegraphics[width=\textwidth]{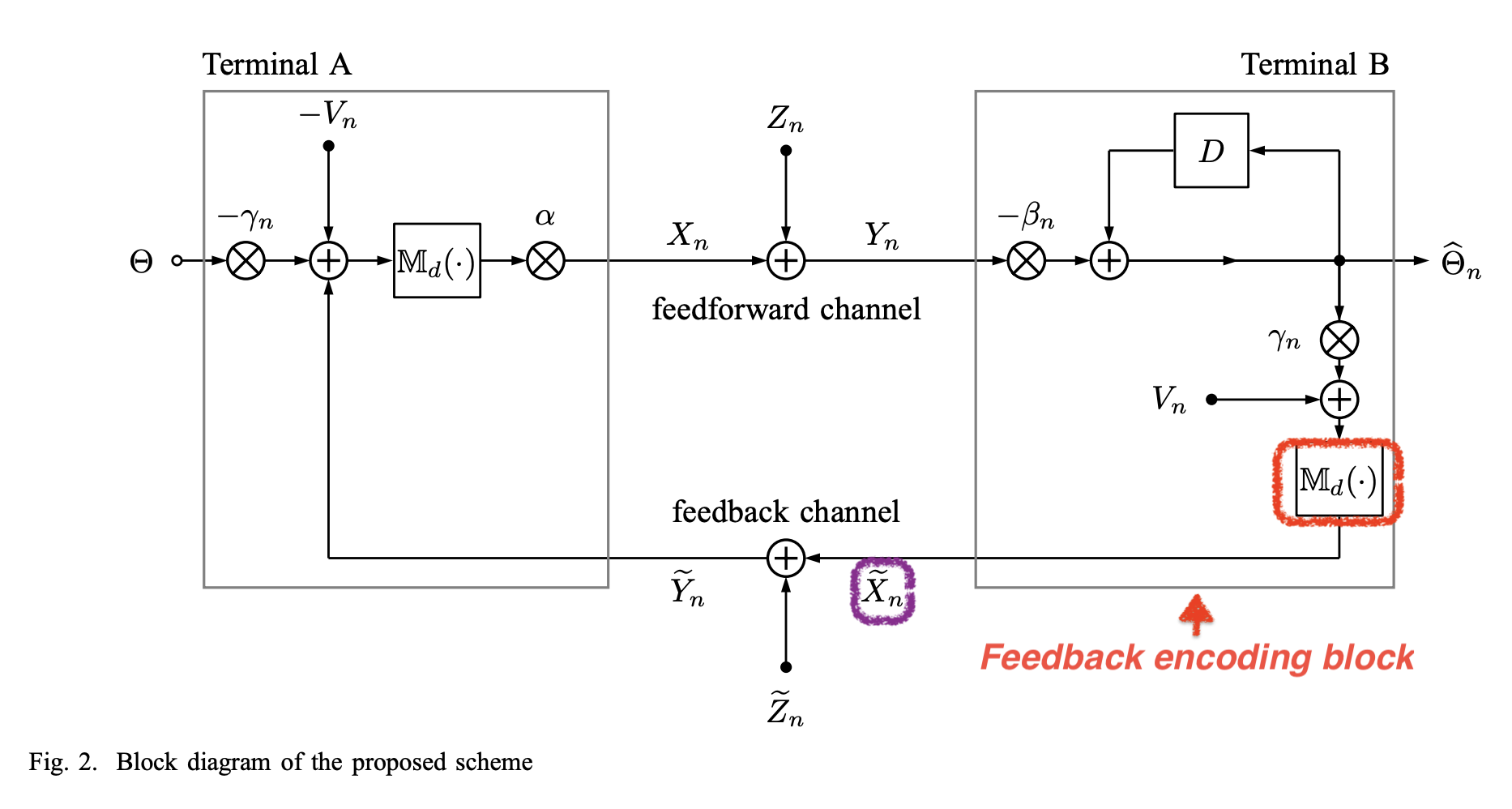}
    \caption{\label{fig:fig2} Figure 2 of the Modulo-SK  paper~\cite{interactive}}
  \end{subfigure}
  \caption{Diagrams showing the difference between  Deepcode that only uses the {\color{black}(possibly delayed)} passive output feedback and Modulo-SK that  requires active feedback {\color{black} with unit delay}.} 
  \label{fig1}
\end{figure}

\section{Neural code for active feedback in~\cite{kim2018deepcode} and Modulo-SK are also designed for, and work in, different settings.}
As noted in the previous section, our main focus is on the passive feedback setting. 
We did discuss  active feedback in one paragraph of the conference version (Section 4 in~\cite{kim2018deepcode}). Even there, the setting is different from that of Figure~\ref{fig1}(b): we study AWGN channels with a $K$-step {\em delayed} active feedback, where $K$ denotes the number of {\em information bits} ($K$=50 is considered in the relevant experiment  in ~\cite{kim2018deepcode}). This is a different (and strictly harder) setting from that in Figure~\ref{fig1}(b) which has unit step delay. 
%
%
Hence, the comparison in Section~2 of \cite{shayevitz2018} between the Modulo-SK scheme and Deepcode is not appropriate.  

Apart from this basic inconsistency, we note two more. 
\begin{enumerate}
\item 
 We have consistently used the terminology ``Deepcode" for a specific scenario: AWGN channel with  unit-step delayed \emph{output} feedback, i.e., the passive feedback scenario. In the case of active and/or delayed feedback we have used the generic terminology of ``neural code" (see legend in  Figure~1(a) of \cite{shayevitz2018}). Following  our nomenclature, the title of \cite{shayevitz2018} is misleading.  

\item 
In  \cite{shayevitz2018}, the authors write that ``Deepcode is .... claimed to be superior to all previous schemes in the literature". 
As we have clarified in the previous section, our claimed superiority of Deepcode relates  only to passive feedback. We have never claimed state-of-the-art performance in the $K$-step delayed active feedback setting.  
%
The main point of the   experiments  in  Section 4 in~\cite{kim2018deepcode} is to ($a$) explore neural codes for practical feedback scenarios (e.g., delayed feedback) and ($b$) demonstrate that the neural network based approach can be extended to allow coding of the feedback signal and achieves an improvement compared to the passive feedback. This experiment is interesting from a machine   learning perspective: it is a very challenging task to jointly learn 3 functions:  channel encoding, feedback encoding, and channel decoding. 
\end{enumerate}
%

 {\bf Summary}. 
\begin{itemize}
    \item Our study in the Deepcode project is focused on output feedback. We discussed active feedback in a short paragraph in the conference version \cite{kim2018deepcode} and eliminated it  in the journal version \cite{deepcode2018jsait}. Further,  the  short   discussion of active feedback in the conference version is when the feedback is {\em  delayed} (in our study the delay is significant -- by as many steps as the number of information bits). The  comparison of the neural code (in~\cite{kim2018deepcode}) and Modulo-SK in~\cite{shayevitz2018} is simply not appropriate because Modulo-SK is designed for, and evaluated with, unit delay  in the feedback.
    \item The neural code compared to Modulo-SK in~\cite{shayevitz2018} is not Deepcode. 
    Deepcode is designed for channels with noisy output feedback; i.e., feedback is the received value itself. 
    \item We have not claimed the neural code we constructed  in the $K$-step delayed active feedback setting is state-of-the-art.  Furthermore, we have noted that ``improving further the performance of
(active) Deepcode at realistic feedback SNRs (such as 10dB or
lower) is an important open problem." (Section VI in \cite{deepcode2018jsait}). 
\end{itemize}

\begin{table}[!ht]
    \centering
    \begin{tabular}{c|c|c}
        Coding scheme & feedback delay & feedback encoding \\
        \hline 
        Deepcode~\cite{kim2018deepcode,deepcode2018jsait} & 1 & No \\
        Modulo-SK~\cite{interactive} & 1 & Yes \\
        Neural code for a $K$-step delayed output feedback~\cite{kim2018deepcode} & K & No \\
        Neural code for a $K$-step delayed active feedback~\cite{kim2018deepcode} & K & Yes \\
    \end{tabular}
    \caption{Comparison of Deepcode, Modulo-SK, and Neural codes for a $K$-step delayed feedback.}
    \label{tab:my_label}
\end{table}

\section{The Precision Issue with SK and Robust Implementations}

 \subsection{Precision issue with SK already discussed in \cite{deepcode2018jsait}} 
The  precision issue with SK (Schalkwijk and Kailath scheme \cite{schalkwijk1966coding}) is  discussed  in Section IV  of the journal version~\cite{deepcode2018jsait}. We make two points here.  First, 
when the feedback is noiseless,  
we already demonstrated that 
the precision issue of SK code can be mitigated by reducing the coding block length. By doing so, SK can outperform Deepcode \emph{only when} the precision is large enough ($>$ 8-bit) and feedback is noiseless (Figure 14 taken from the journal version~\cite{deepcode2018jsait} of Deepcode). Second,  we note that Deepcode outperforms SK regardless of the encoding block length of SK for noisy feedback settings.



\begin{figure}[!ht]
\centering
\includegraphics[width=0.7\textwidth]{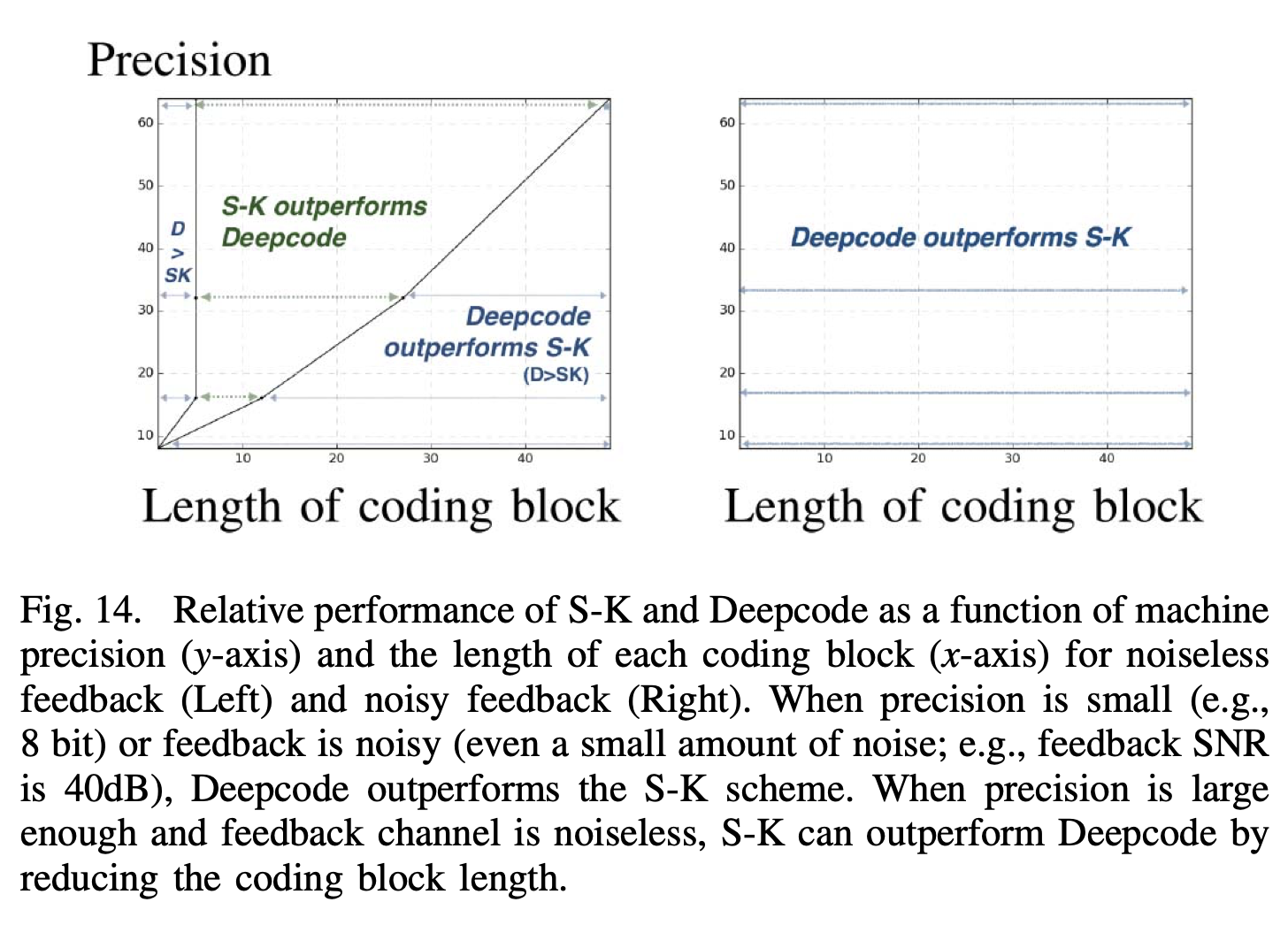}
\caption{Figure 14 from the journal version~\cite{deepcode2018jsait} of Deepcode illustrating whether Deepcode outperforms SK or not as a function of precision (8, 16, 32, 64-bits) and the length of coding block for SK (1 to 49).  
}
\label{fig2}
\end{figure}

\subsection{Implementation of SK}
\begin{figure}[!ht]
    \centering
    \includegraphics[width=0.5\textwidth]{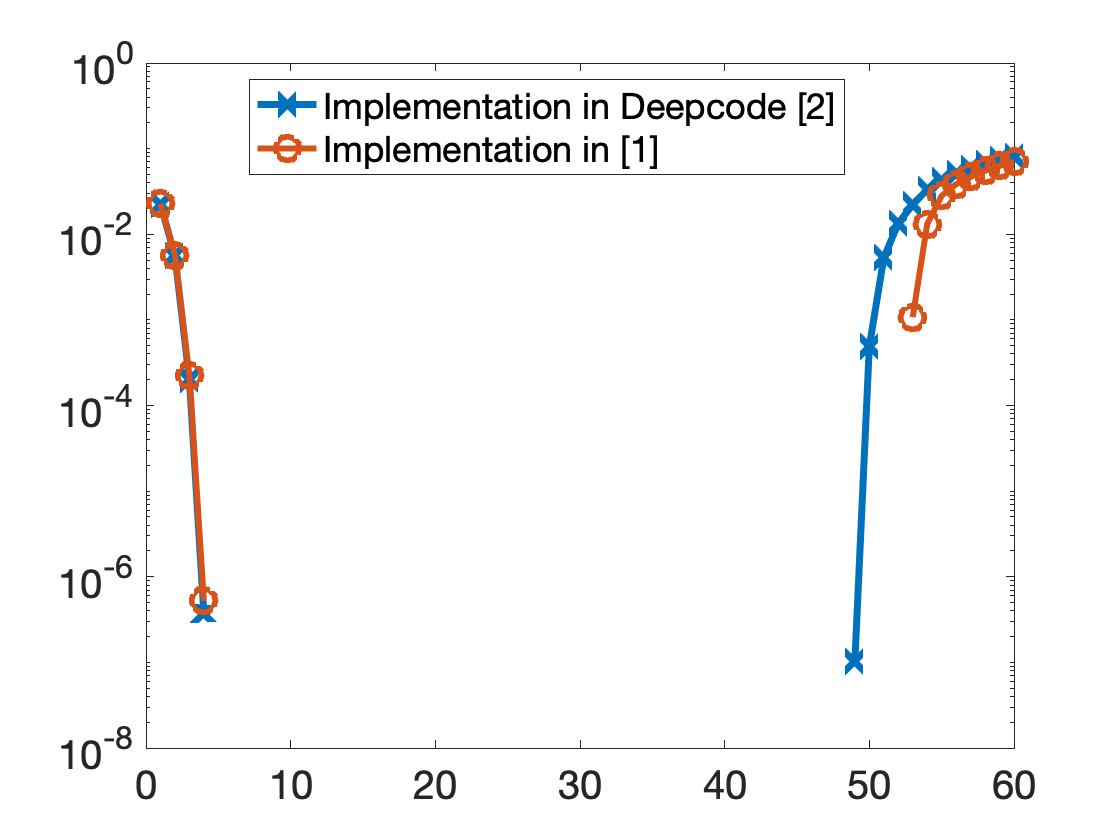}
        \put(-252,100){BER}
	\put(-170,-13){Coding block length $K$}

    \caption{
    The claimed superiority of the implementation of SK scheme from \cite{shayevitz2018} is marginal. BER as a function of coding block length $K$ simulated via~\cite{shayevitz2018} from \url{https://github.com/assafbster/Modulo-SK} (-o-) on AWGN channels with 0dB feedforward SNR and noiseless feedback with the native MATLAB double (64-bit) precision. Precision issue arises as coding block length increases (K $\geq 53$) in the implementation of~\cite{shayevitz2018}. As a reference, BER from the implementation in~\cite{kim2018deepcode} (-x-) is also shown. Precision issue arises as coding block length increases (K $\geq 49$) in the implementation of~\cite{kim2018deepcode}.  }
    \label{fig:prec}
\end{figure}

In~\cite{shayevitz2018}, the authors  claim that ``SK can in fact be implemented in a numerically stable way for any number of information bits and iterations, by a judicious fixed point or floating point implementation that takes into account the required resolution of the signals in the system" and provide a corresponding implementation of SK.  We execute this  implementation  and find that this claim is not true. In Figure~\ref{fig:prec}, we plot 
the BER as a function of the coding length $K$ (coding rate is fixed at 1/3) for 0dB forward channel with noiseless feedback for the implementation in~\cite{shayevitz2018} from \url{https://github.com/assafbster/Modulo-SK} (-o-). 
For $K \ge 53$, the BER of~\cite{shayevitz2018} starts to increase drastically. The BER is averaged over $10^7$ runs; for $5 \leq K \leq 52$, the BER is below $10^{-8}$. We reiterate that the BER of Deepcode remains unchanged with varying $K$.

 In Figure~\ref{fig:prec}, we also plot the BER of our implementation in~\cite{kim2018deepcode} (-x-) as a reference, for which BER starts to increase after $K \ge 49$. Both  SK implementations, ours and the implementation of~\cite{shayevitz2018}, achieve  similar BERs for small values of $K$, but start to deviate as $K$ increases. We emphasize that both implementations suffer from increasing precision errors as the coding block length $K$ increases, although the critical values of $K$ (at which the error increases drastically) differ, although the difference is small. The key point, though, is that DeepCode is vastly superior to both SK implementations, in this regime. 

\begin{figure}[!ht]
    \centering
    \includegraphics[width=0.6\textwidth]{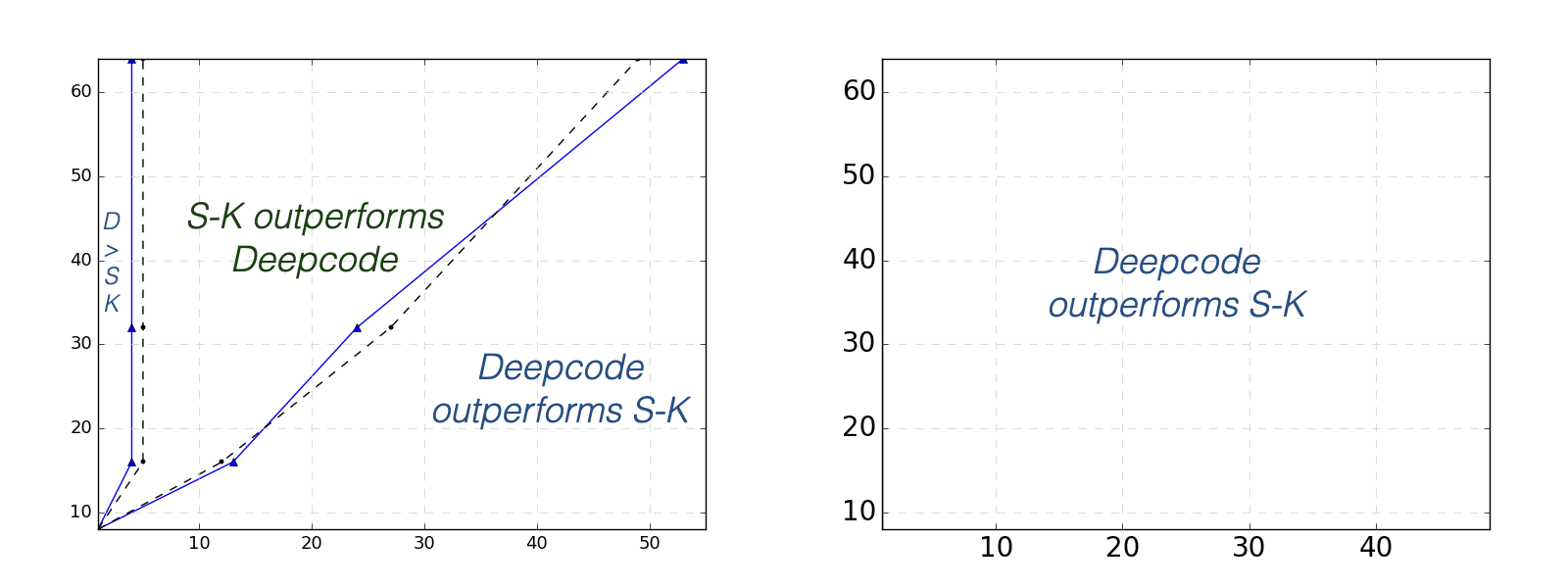}
    \put(-304,103){Precision}
	\put(-260,-13){Coding block length}
	\put(-260,-26){(\emph{Noiseless feedback})}
    \put(-152,103){Precision}
	\put(-120,-13){Coding block length}
    \put(-120,-26){(\emph{Noisy feedback})}
\includegraphics[width=0.3\textwidth]{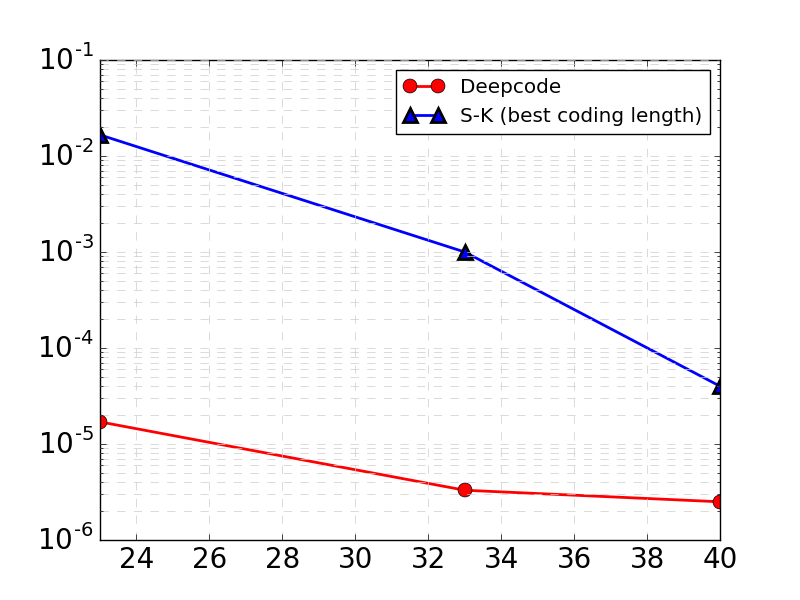}
    \put(-152,103){BER}
	\put(-110,-13){Feedback SNR (dB)}
	
    \caption{(Left): Fig.~\ref{fig2} (left) re-generated with the SK implementation in~\cite{shayevitz2018} from \url{https://github.com/assafbster/Modulo-SK} (solid line). 
    As a comparison, Fig.~\ref{fig2} generated with the SK  implementation in~\cite{kim2018deepcode} is shown in dotted lines. The results remain unchanged except for a slight shift in the transition boundaries. (Middle) Fig.~\ref{fig2} (right) re-generated with the SK implementation based on~\cite{shayevitz2018}. For noisy feedback channels, Deepcode outperforms SK implementation in~\cite{shayevitz2018}  regardless of the coding length. 
    (Right): BER vs feedback SNR, where forward SNR is fixed as 0dB. Deepcode outperforms the SK code with the optimized coding block length, implemented based on~\cite{shayevitz2018} and tested with the default MATLAB 64-bit precision. The best coding length is $K=2,3,4$ for 23dB, 33dB, 40dB feedback, respectively for the SK with the best coding length (-$\bigtriangleup$-). 
    For Deepcode (-$\circ$-), information size of $K=50$ is used throughout.}
    \label{fig:new}
\end{figure}
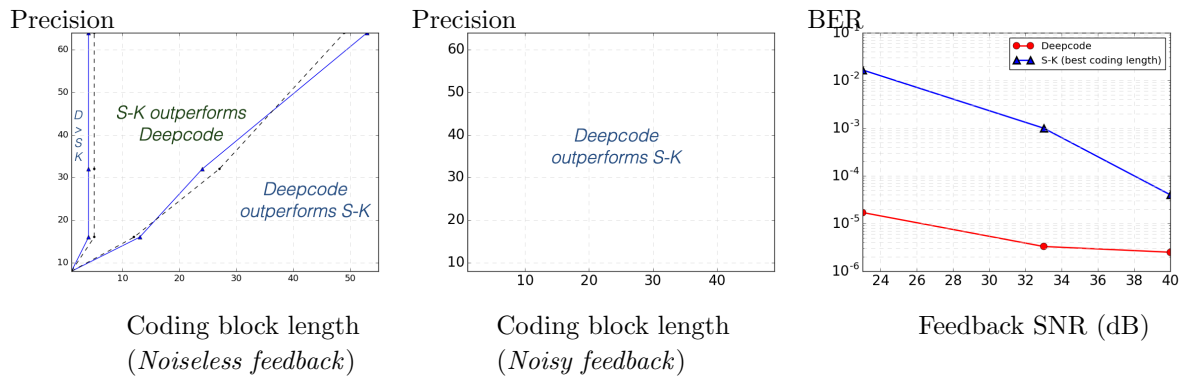

There are two main differences between the two implementations: (i) power allocation. We optimized the power allocated to the first transmission vs. the power allocated to the rest of transmissions as noted in~\cite{gallager} while the uniform power is allocated throughout in~\cite{shayevitz2018}; (ii) 
the SK scheme can be written in two different ways (identical in the infinite precision case) and~\cite{shayevitz2018} and~\cite{kim2018deepcode} implement different representations. We empirically found the key difference on the precision break point (K=53 vs K=49) arises from the latter, which we elaborate in 
Algorithm 1. We re-run our experiment in~\cite{kim2018deepcode} with the implementation of~\cite{shayevitz2018} and verify the claims in the~\cite{kim2018deepcode} remain essentially unchanged, as shown in Figure~\ref{fig:new} (left, middle). 

{\color{black} It is important to note that for noisy feedback settings, Deepcode outperforms SK regardless of the coding block length used for SK (Figure~\ref{fig:new} (middle)). 
To further clarify on what happens for noisy channels, 
in Figure~\ref{fig:new} (Right), we compare the SK (with best coding length) and Deepcode for varying feedback SNR: for the SK, we empirically find the best coding block length using the SK implementation based on~\cite{shayevitz2018}\footnote{We modified the implementaion in~\cite{shayevitz2018} to incorporate noise added to the transmission of output (received values) in the feedback channel.} and plot the best BER. As one might envision, reducing the coding block lengths helps in improving the reliability; the best coding block length $K$ turns out to be $2,3,4$ for feedback SNRs 23dB, 33dB, and 40dB, respectively. Nevertheless, DeepCode is more reliable than the optimized BER of the SK. }




\begin{algorithm}[!ht]
 \caption{Shalkwijk-Kailath Implementations in~\cite{shayevitz2018} and~\cite{kim2018deepcode}}
\SetAlgoLined
 Input $\Theta$: a (normalized) $2^K$-ary PAM symbol;\\
 $U_0 = \Theta$; \\
 $X_0 = U_0$;\\
 $Y_0 = {\rm AWGNchan}(X_0,SNR)$;\\
 $\hat{\Theta}_0 = Y_0$;\\
 \While{n = 1, $\cdots$, N}{
\uIf{Implementation in~\cite{shayevitz2018}}
    {$U_n = \hat{\Theta}_{n-1} - \Theta$; \\
  }
  \uElseIf{Implementation in~\cite{kim2018deepcode}}
    {\uIf{n == 1}{
    $U_1 = Y_0 - X_0$;\\
    }\Else{$U_n = U_{n-1} - E[U_{n-1}|Y_{n-1}];$ \\}
     }
  $\alpha_{n}$: power normalization constant (updated based on $\alpha_{n-1}$ and SNR)\\
  $X_n = \alpha_n U_n$; \\
  $Y_n = {\rm AWGNchan}(X_n,SNR)$; \\
  $\hat{\Theta}_n = \hat{\Theta}_{n-1} - E[U_n|Y_n] $\\
 }
\end{algorithm}

\bibliographystyle{unsrt}
\bibliography{ref.bib}

\begin{thebibliography}{1}

\bibitem{shayevitz2018}
Assaf Ben-Yishai and Ofer Shayevitz.
\newblock Simple modulo can significantly outperform deep learning-based
  deepcode.
\newblock {\em arXiv:2008.01686v2}, August 2020.

\bibitem{kim2018deepcode}
Hyeji Kim, Yihan Jiang, Sreeram Kannan, Sewoong Oh, and Pramod Viswanath.
\newblock Deepcode: Feedback codes via deep learning.
\newblock In {\em Advances in Neural Information Processing Systems}, pages
  9436--9446, 2018.

\bibitem{schalkwijk1966coding}
J~Schalkwijk and Thomas Kailath.
\newblock A coding scheme for additive noise channels with feedback--i: No
  bandwidth constraint.
\newblock {\em IEEE Transactions on Information Theory}, 12(2):172--182, 1966.

\bibitem{kim2007gaussian}
Young-Han Kim, Amos Lapidoth, and Tsachy Weissman.
\newblock The gaussian channel with noisy feedback.
\newblock In {\em 2007 IEEE International Symposium on Information Theory},
  pages 1416--1420. IEEE, 2007.

\bibitem{kim2011error}
Young-Han Kim, Amos Lapidoth, and Tsachy Weissman.
\newblock Error exponents for the gaussian channel with active noisy feedback.
\newblock {\em IEEE Transactions on Information Theory}, 57(3):1223--1236,
  2011.

\bibitem{deepcode2018jsait}
H.~{Kim}, Y.~{Jiang}, S.~{Kannan}, S.~{Oh}, and P.~{Viswanath}.
\newblock Deepcode: Feedback codes via deep learning.
\newblock {\em IEEE Journal on Selected Areas in Information Theory},
  1(1):194--206, 2020.

\bibitem{interactive}
A.~{Ben-Yishai} and O.~{Shayevitz}.
\newblock Interactive schemes for the awgn channel with noisy feedback.
\newblock {\em IEEE Transactions on Information Theory}, 63(4):2409--2427,
  2017.

\bibitem{gallager}
R.~G. Gallager and B.~Nakiboglu.
\newblock Variations on a theme by {S}chalkwijk and {K}ailath.
\newblock {\em IEEE Transactions on Information Theory}, 56(1):6--17, Jan 2010.

\end{thebibliography}

\end{document}